# Fabrication of nanopores in a graphene sheet with heavy ions: a molecular dynamics study

Weisen Li[1], Li Liang[1], Shijun Zhao[2], Shuo zhang[1], Jianming Xue[1,2*]

[1] State Key Laboratory of Nuclear Physics and Technology, School of Physics, Peking University, Beijing 100871, P. R. China

[2] Center for Applied Physics and Technology, Peking University, Beijing 100871, P. R. China

**Abstract**

Molecular dynamics (MD) simulations were performed to study the formation process of nanopores in a suspended graphene sheet irradiated by using energetic ions though a mask. By controlling the ion parameters including mass, energy and incident angle, different kinds of topography were observed in the graphene sheet. Net-like defective strucutures with carbon atom chains can be formed at low ion fluence, which provides the possibility to functionalize the irradiated sample with subsequent chemical methods; finally a perfect nanopore with smooth edge appears as the ion fluence is high enough. We found that the dependence of ion damage efficiency on ion fluence, energy and incident angle are different from that predicted by the semi-empirical model based on the binary-collision approximation, which results from the special structure of graphene. Our results demonstrate that it is feasible to fabricate controlled nanopores/nanostructures in graphene via heavy ion irradiation.

**Key words:** Graphene; Ion Irradiation; Nanopore; MD Simulation

**PACS code:** 61.48.Gh; 79.20.Rf.

# 1. Introduction

Graphene is a monoatomic layer of carbon atoms strictly arranged into a honeycomb lattice, and has become one of the most studied materials due to its excellent properties and promising applications in novel nano-devices[1-3]. It has been pointed out that the nanopore created in a graphene sheet is the most prospective candidate for DNA sequencing with current-blocking method because of its ultrathin thickness that enables one to distinguish single base[4, 5]. Recently graphene with nano-holes found its new applications such as in sea water desalination[6, 7]. How to drill nanometer pores in a graphene sheet becomes an urging and interesting topic.

Currently nanometer pores in graphene sheets are usually generated by electron beam irradiation [8]. Compared with electron beam, heavy ions are more efficient to remove atoms from graphene because of its mass is at least 3 orders larger than electron, which means heavy ions can transfer more kinetic energy to the target atom. Energetic ions has been demonstrated to be able to cut and create nano-structures by removing part of carbon atoms in graphene sheet[9]. Anyhow, not like electron beams, ion beam cannot be focused into such a small spot to fabricate nano-scale pores[10]. To drill nanopores in graphene with energetic ions, cluster ions containing thousands of atoms have been proposed to be used in irradiation [11, 12]. Anyhow, it is technically difficult to produce such cluster ions to irradiate graphene and no experimental work has been performed so far. A feasible way is to use the ion track lithography method, which allows the energetic ions pass through nanoholes in a mask and then bombard the graphene sheet only within a nano-scale confined area. This method has been used successfully to fabricate nanostructures[13]. It is interesting and helpful for the future experiment to predict what kinds of nano-structures could be produced in the suspended graphene sheet by using this lithography technique, and how to achieve desirable defective structures by controlling ion parameters.

The irradiation damage introduced by incident ions in bulk materials can be calculated with high degree of accuracy by a semi-empirical method based on the binary-collision (BC) approximation, combined with statistical algorithms to account for how a moving ion transfers its energy to the sample atoms[14, 15]. However, it has been pointed out that the theory of irradiation effects for bulk targets does not always work for the low-dimensional materials[16, 17]. In particular, it is recently demonstrated that the BC approach is not applicable to graphene, and an explicit account of the atomic structure is required for the atomically thin target of graphene[18, 19]. In these previous works, single ions of different parameters were used to bombard graphene sheets repeatedly and the results are the average of these simulations[18-20], hence their results cannot be used to estimate the damage process in ion track lithography, in which successive incident ions are used experimentally. For the purpose of creating nanopores in graphene, it is crucial important to know the evaluation of defective structures under successive ion irradiation at different parameters, which has not been performed so far.

It is well known that energetic ions damage the materials via two ways: the nuclear collision and the electronic excitation. Due to its excellent electronic and thermal conductance, most of the time, the electronic excitation process could be neglected in graphene and the nuclear collision dominates the damage process especially during ion irradiations with not very high energies. This is different from bulk materials as the deposition of energy into electronic degrees of freedom occurs under impacts of energetic ions[21]. For this reason, we simulated impacts of energetic ions with suspended graphene sheet using classical molecular dynamics simulation by taking into account only the nuclear collisions between the moving ions and the target carbon atoms.

In this paper, we present our MD simulation results on the formation process of nano-scale defective structures in graphene by continuous incidence of energetic ions. The influences of ion mass, energy and incidence angle on the damage efficiency will be discussed. Besides, the defects in graphene induced by irradiation are also studied, especially the carbon chains and the dangling carbon bonds at the edge of the nanopore, which will act as reactive centers for the chemical functionalization in a subsequent properties modification .

## 2. Simulation details

In our MD simulation, the size of target graphene system is approximately 30 nm×30 nm. The temperature of atoms at four borders was kept at 300 K with a Berendsen thermostat[22] . C, Si and Au ions were used to simulate the ion impacts, and the maximum number of each kind of incident ions was always 300. These ions bombarded a graphene sheet continuously and randomly inside a circle with radius of 1.5 nm as shown in Fig. 1, which is to mimic the confined irradiation area by energetic ions though a mask with cylindrical nanoholes. Energy of incident ions was ranged from 500 eV to 100 keV, and the incident angle was taken from $0^o$ (normal direction) to 75° with respect to the normal line. It should be pointed out that in previous work[18] , no matter how many times the incidence event is, the graphene sample always keeps as a perfect structure before the irradiation of the next ion, in other words, the sample is restored to its pristine formation after the last irradiation ion. This is necessary for studying the primary damaging efficiency under ion irradiation, like sputtering yield, vacancy production probability *et al*., but it is not suitable for simulating the process of fabricating a nano-size defect structure, like nanopores in the graphene, which needs a continuous irradiation of energetic ions. Therefore, in our simulation, the damaged structure is remained during the irradiation of next ions, and the final damage is the accumulation of all the incident ions.

The atomic interaction among the carbon atoms in graphene is described with AI−REBO potential [23], and the interaction of the incident C, Si ions with the target carbon atoms is modeled by the tersoff/zbl potential including a 3−body Tersoff potential and the Ziegler−Biersack−Littmark (ZBL) universal repulsive potential[24, 25]. This hybrid potential can correctly describe the ion-carbon interaction compared to the pure ZBL potential that is used widely in conventional BC model[26], which is in particular important for ions that can form chemical bonds with carbon. The incident Au ions and target atoms are calculated only with the pairwise ZBL potential because of not chemical bond can be formed between Au and carbon. After each ion impacts, the system is allowed to relax for 1 picosecond to make sure that the system has reached a local energy minimum state, while longer relaxation time makes no change to the final state.

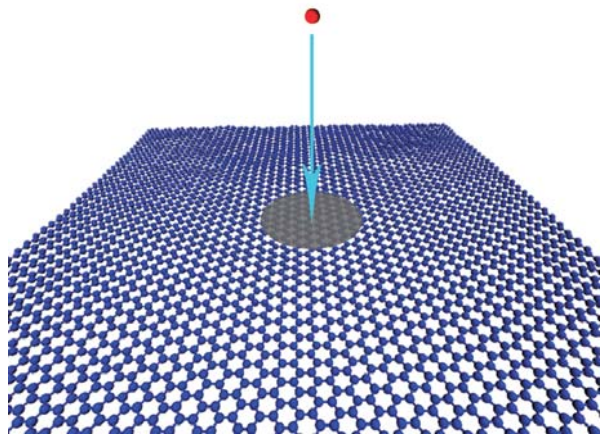

Fig. 1. Sketch of the simulation model. The gray circle region in the center represents a ion incident zone, which means all ions will bombard the graphene inside this circle continuously and randomly, whose radius is 1.5 nm.

## 3. Results and discussion

### 3.1  Defective nano-structures formed by ions of different parameters

We observed the topography of graphene sheet during ion irradiation with increasing numbers. In the simulation, C, Si and Au ions with energy of 1 keV are used to irradiate the graphene one by one continuously. The topography of the irradiated graphene samples as function of ion numbers are shown in Fig .2, only the results for ion numbers of 50, 100, 150, 200, 250 and 300 are shown. From top to bottom: C, Si and Au, respectively. Since we focus on the large defective structures such as nano-holes formed by ion irradiation, we do not show our results with ion number less than 50, which usually forms simple defects that had been investigated in previous works.

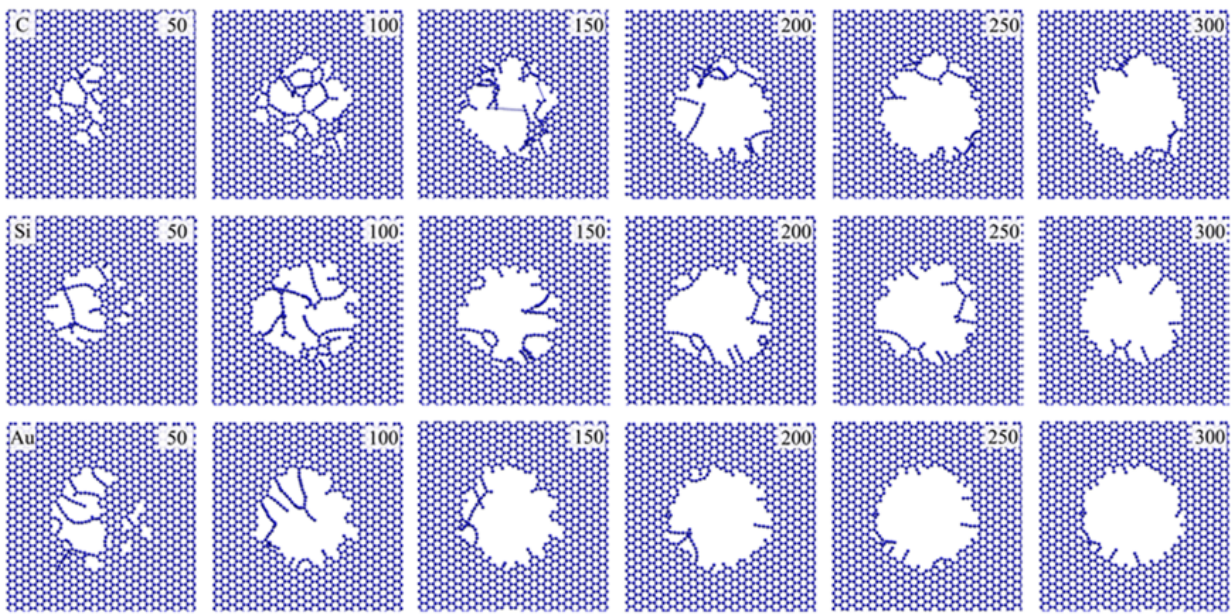

Fig. 2. Topography of the irradiated region  by ions with different masses and different incident numbers. From top to bottom  the three rows were drilled by C, Si, and Au ions respectively. From left to right in each row the number of incident ions is 50, 100, 150, 200, 250 and 300. All the incident ion energy is 1 keV.

It can be seen from Fig.2, the irradiated region shows different characteristics resulting from more and more carbon atoms are gradually removed as the ion number increases. With 50 C ions irradiation, some small holes appear in the graphene sheet, but these defective structures are discrete. With continuous bombardment of ions, these small holes get connected and grow larger, finally a whole nanopore with many carbon chains on the pore edge is formed after being irradiated with 200 ions. Further increasing ion number to 300, the carbon chains become short and the pore edge becomes smooth.

One interesting phenomenon is worthy of mentioning, some carbon chains appear during the irradiation, and some short carbon chains still exist near the pore edge even when the ion number is as high as 300. Though our primary purpose is to drill nanopores in graphene sheet, the irradiation induced defects observed in our simulation especially the carbon chains are also very useful in functionalizing graphene with chemical methods. In general, it is quite difficult to decorate the graphene chemically with bio-molecules because of its inert chemical property. However, after ion irradiation this will be much easier due to the formation of these carbon chains, which provide a high concentration of dangling carbon bonds, and chemical decoration can be achieved on these reactive sites [27].

For Si and Au ions, the evaluation of the defective structures is similar to C ions, but they are more efficient to remove carbon atoms from the graphene, which means less ions are needed to form same defective structure than carbon ions. This behavior results from the fact that heavy ions can transfer more kinetic energy to graphene atoms in the collision compared to light ions with same energy. Details of the underlying mechanism will be discussed in the next section.

The evaluation of the defective structures is also greatly influenced by the ion energy, as shown in Fig.3. By fixing the ion number at 300, we compare the final topography of greaphene sample bombarded with ions at five energies: 500 eV, 1 keV, 5 keV, 50 keV and 100 keV. The results indicate that, a higher energy means a lower damage efficiency. For C ions with energies higher than 5 keV, a complete nanopore cannot be formed even when ion number is as high as 300. Comparing the three rows in Fig.3, the influence of this energy dependence gradually vanishes as the ion mass increases. For the case of Au, through the whole energy range used in our simulation, a good nanopore profile can be always obtained; what's more, an incident ion energy of 5 keV created the most smooth nanopore with fewest carbon chains on the pore edge.

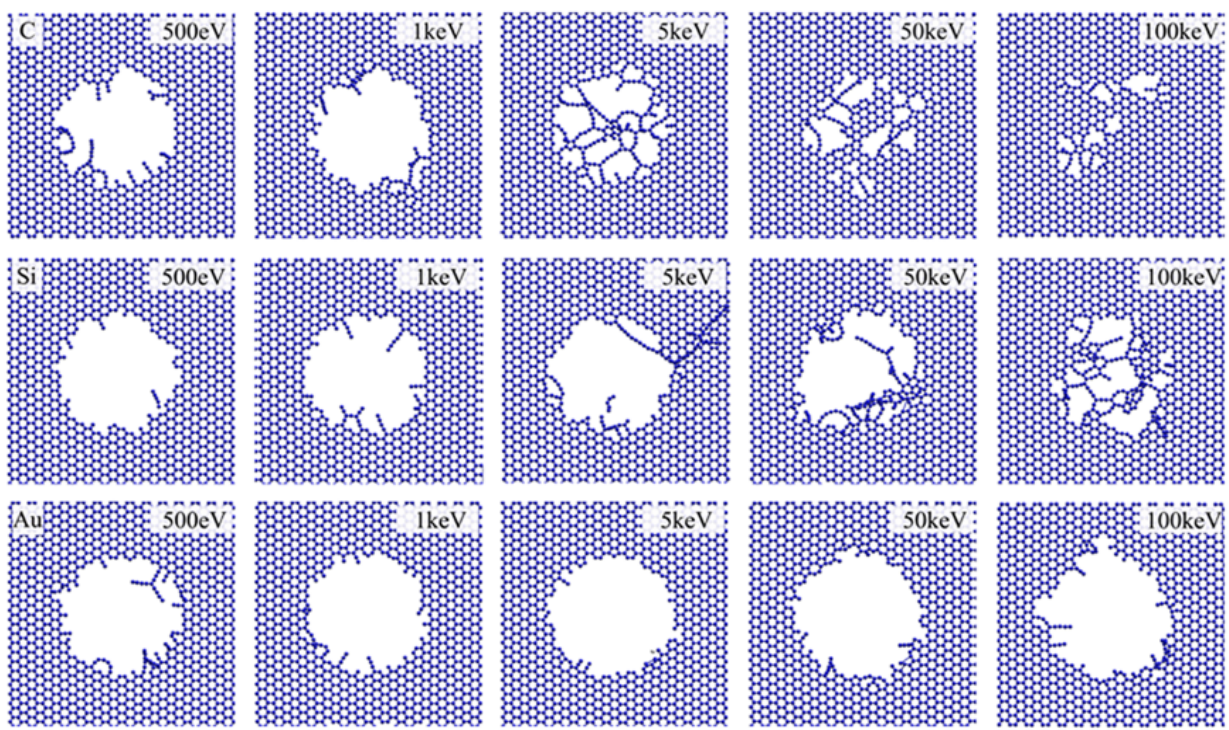

Fig. 3. Comparison of nanopores fabricated by different incident ions with different incident energies at ion number of 300.. Nanopores presented in the above three rows were drilled by C, Si, and Au ions from top to bottom respectively. The incident ion energy from left to right in every row is 500 eV, 1 keV, 5 keV, 50 keV and 100 keV.

Various defective structures could be formed by ion induced irradiation damage, and the damage efficiency depends on ion number, mass and energy. Heavy ions with low energy are most efficient in drilling uniformly shaped pores in experiments. Besides formation of perfect nanopores, by carefully choosing ion parameter one can obtain desirable defective nano-strcuctures in graphene sheet, such as dangling carbon chains in purpose of chemically decoration.

## 3.2  Dependence of damage efficiency in graphene by continues ion beams

The above qualitative analysis indicates that the efficiency for different ions to remove carbon atoms depends on the parameters of incident ions. To quantitatively understanding the underlying mechanism we calculate the numbers of sputtered carbon atoms at different numbers of incident ions. The sputtered atom is defined as the atom with a distance longer than 1.5 nm to the graphene, beyond which the atom does not have potential connection with the graphene atoms any longer. The sputtering yield is defined as the average sputtered atoms per incident ion.

We analyze the sputtered atom numbers and the statistic results are shown in Fig. 4 as functions of ion mass and ion number, in Fig. 5 as a function of ion energy, and in Fig. 6 as a function of ion incident angle.

Different from the case in bulk, in which the sputtering yield is almost a constant during the whole collision, Fig. 4 shows that the dependence of sputtering yield on the number of incident ions in suspended graphene has three stages: When the sputtered atom number is less than around 70, it increases almost linearly with ion number; after that, the yields experience a faster growth until the sputtered atom number reaches ~160; beyond this value, the increase of sputtered atoms number becomes very slowly and nearly saturated. These three stage is very much the case with heavy ions Au, Si and C, though the ion number for these ions to enter the three stages are different.

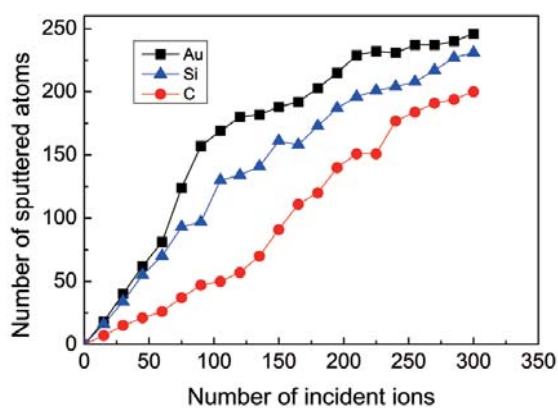

Fig. 4 Variation of the numbers of displaced carbon atoms from graphene sheet as a function of incident ion number for different ion masses. For all these simulations, the ion energy is 1 keV.

This special tendency is due to the single layer structure of graphene. In the first stage, the graphene is not much destroyed by the ion irradiation. In this stage, the sputtered atoms are single atoms and its number increases linearly with the incident ions number; In the following second stage, the graphene has been partly damaged, it is possible that one incident ion could lead to the removal of some atoms as a cluster or as a carbon chain, which results in a much faster increase of the sputtered atoms number. In this stage, some ions may miss colliding with carbon atoms and just pass through the holes formed by previous ions, but this effect is overwhelmed by the cluster/chain effect, so that the number of sputtered atoms shows a faster increase; Finally, most of the atoms in the irradiated region have been sputtered, the probability for the incident ions colliding with the remained few atoms are quite low, therefore the sputtered atom rate becomes very slow, namely the stage 3 as shown in Fig.4.

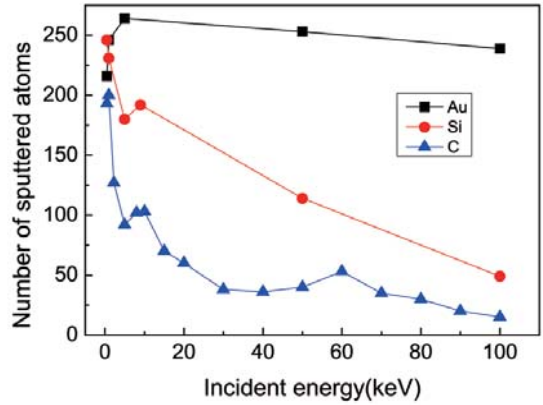

Fig. 5 Number of displaced carbon atoms from graphene sheet as a function of incident ion energy, and the number of incident ions is fixed at 300.

Besides ion mass, ion energy also have significantly influence on the damage efficiency. A big difference is observed in the yield of sputtered carbon atoms as the ion energy increases from 500 eV to 100 keV for different ions in Fig. 5. Heavy ions produce a much higher sputtering yield in the whole energy range, which means it is more efficient in drilling the hole. Another very interesting thing is the trend of sputtering yield. As the sputtering process is mainly dominated by the nuclear energy transfer process, this tendency shows that heavy ion has a much duller energy sensibility of collision cross section than light ion does, and this is much similar for the nuclear stopping power($S_n$) defined in a energy transfer process happening in the bulk material, although this definition is no longer valid in one atom layer structured graphene.

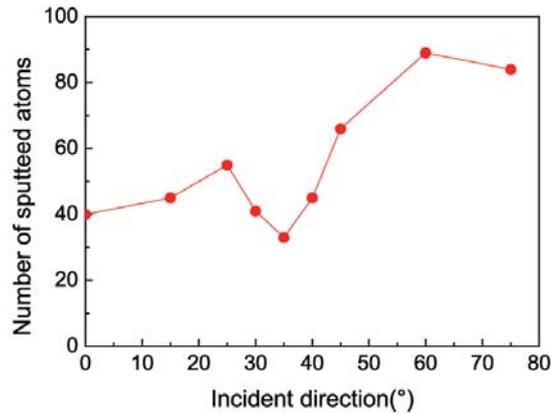

Fig. 6 Number of displaced carbon atoms from graphene sheet as a function of incident ion direction in the case of C ions irradiation with an energy of 50 keV..

The incident angle of ions also has an influence on the sputtering efficiency. The number of displaced atoms as a function of incident angle caused by 300 carbon ions with energy of 50 keV is shown in Fig. 6. In general, the sputtering efficiency keeps a low growth rate as the incident angle increases in the whole range, except for a drop around 35º. The phenomenon is interpreted in two ways. On one hand, a larger angle means a larger effective thickness of the graphene layer and more kinetic energy will be transferred to the graphene sample, therefore it could be expected that the sputtering efficiency also increases; on the other hand, the minimum energy for the graphene atom to escape is different with the emitting angle, and a higher energy is required at some special angles[27] . For ions used in our simulation, this angle is around 30º ~ 40º, which results in the drops of the sputtering efficiency around incident angle of 35º as shown in Fig. 6.

One thing is worthy of mentioning is that, our simulation results with suspended graphene sheet cannot be compared with experimental data so far, which are obtained by bombarding graphene on some kinds of substrate[28-31]. This is because, the damage of the graphene on substrate is greatly influenced by the substrate but not only the incident ions[32], this effect usually will improve the damage efficiency of energetic ions, and the defective structure in the irradiated area will also be totally different.

In general, an ion with higher nuclear stopping power calculated with amorphous carbon can produce more sputtering atoms, and is more efficiency in drilling nanoholes in graphene. By controlling ion parameters, the characteristics of the irradiation region can also be adjusted for different applications. Ions irradiation parameter used in our simulation is equal to an equivalent ion fluence of $4.24 \times 10^{15}$ ions/cm$^2$ in experiments for the case of 300 incident ions in our confined regions, and this ion fluence can be easily reached by most ion accelerator facilities, which provides us the possible way to test our model in an experiment.

**4. Summary**

In summary, we investigate the formation of nanopores in the suspended graphene with different energetic ions. Our results suggest that graphene under impacts of Au can form a relatively perfect nanopore with smooth edges in the irradiated region, while it is not recommended using light ions Si or C under the same conditions. Under the impact of incident C ions, a nanopore can be formed but its edge is irregular, accomplished by many undesired carbon chains. However, this irregular edges can be used as a chemical functionalization sites for special purposes.

An experiment on this topic is under investigation in our group, and a subsequent property modification based on these defective nano-structures has also been under discussion. Our simulation work first indicates that heavy ion irradiation is a feasible and convenient way to drill a nano-size hole in graphene, and by a carefully choosing of irradiation ion parameters, a desired and smooth nanopore could be obtained.

**Acknowledgement**

This work is financially supported by the Ministry of Science and Technology of China (Grant No. 2010CB832904) and by NSAF (Grant No. U1230111) and NSFC (Grant No. 91226202).